\documentclass[11pt,a4paper]{article}

\usepackage{latexsym}
\usepackage{amssymb}
\usepackage{amsmath}
\usepackage{amsfonts}
\usepackage{mathrsfs}
\usepackage{cite}
\usepackage{color}
\usepackage{bm}

\catcode`@=11
\renewcommand{\section}{\@startsection{section}{1}{0pt}{\medskipamount}
    {\medskipamount}{\large\bf}} \numberwithin{equation}{section}
\catcode`@=12

\newcommand{\be}{\begin{equation}}
\newcommand{\ee}{\end{equation}}

\def\tr{{\rm tr}}
\def\Tr{{\rm Tr}}
\def\cN{{\cal N}}

\def\bea{\begin{eqnarray}}
\def\eea{\end{eqnarray}}
\def\nn{\nonumber}

\def\cN{{\cal N}}

\def\f{\frac}

\def\nn{\nonumber}

\def\d{\delta}

\def\g{\gamma}

\def\ve{\varepsilon}

\def\sB{\stackrel{\frown}{\square}}

\def\eq{\eqref}
\def\pr{\partial}
\def\nb{\nabla}

\numberwithin{equation}{section}

\begin{document}

\begin{titlepage}

\begin{center}
\vspace{1cm}

{\Large \bf One-loop divergences of effective action in $6D,\, {\cal
N}=(1,0)$ supersymmetric four-derivative gauge theory}

\vspace{0.7cm} {\bf
 A.S. Budekhina\footnote{budekhina@tspu.edu.ru}$^{\,a,b}$,
 B.S. Merzlikin\footnote{merzlikin@tspu.edu.ru}$^{\,d,e}$ }
 \vspace{0.7cm}

 $^a$ {\it Bogoliubov Laboratory of Theoretical Physics, JINR, 141980
 Dubna, Moscow region, Russia \\ \vskip 0.1cm
 $^b$ Department of Theoretical Physics, Tomsk State Pedagogical University,\\
 634061, Tomsk,  Russia \\ \vskip 0.15cm
$^d$ Laboratory of Applied Mathematics and Theoretical Physics, \\
Tomsk State University of
 Control Systems and Radioelectronics, 634050, Tomsk, Russia\\ \vskip 0.1cm
 $^e$ Department of Mathematics and Mathematical Physics, \\
 Tomsk Polytechnic University, 634050, Tomsk, Russia\\ \vskip 0.1cm
}
\end{center}

\vspace{0.1cm}

\begin{abstract}
We consider six-dimensional higher-derivative $\cN=(1,0)$
supersymmetric gauge theory coupled with the hypermultiplet. We use
the background superfield method in six-dimensional $\cN=(1,0)$
harmonic superspace to study the effective action in the theory.
Using the dimensional regularization scheme we analyze the one-loop
divergent contributions to the effective action. We demonstrate that
UV behaviour is determined by the higher-derivative term for gauge
multiplet sector.
\end{abstract}

\end{titlepage}

\setcounter{footnote}{0} \setcounter{page}{1}

\section*{Introduction}
In the present work, we continue the study of structure one-loop
divergences in six-dimensional gauge theories with high derivatives
started in \cite{Buchbinder:2020tnc,Buchbinder:2020ovf}. We consider
the $\cN=(1,0)$ supersymmetric Yang-Mills (SYM) theory interacting
with hypermultiplet in arbitrary representation of the gauge group.
We discuss in details models with up to four space-time derivative
in kinetic terms for component fields in both gauge and
hypermultiplet multiplets. We study the theory using the explicitly
supersymmetric and covariant approach based on the background field
method in $6D, \cN = (1,0)$ harmonic superspace
\cite{Buchbinder:1997ya,Buchbinder:1998np,Galperin:1984av,Galperin:2001uw,Howe:1985ar,Zupnik:1986da}.
We are interesting in the one-loop divergences of the effective
action in the gauge multiplet sector. We demonstrate that divergent
contributions from the quantum gauge multiplet is determined only by
high-derivative term in the classical action and the presence of the
standard two-derivative supersymmetric Yang-Mills kinetic term do
not impact on the UV behavior. The divergent contributions from
quantum hypermultiplet appears from the two-derivative kinetic term
only.

As was mentioned above, we consider six-dimensional $\cN = (1,0)$
supersymmetric gauge theory with four higher derivatives, which was
constructed in \cite{Ivanov:2005qf} using harmonic superspace
formulation. Conventional $\cN = (1,0)$ supersymmetric Yang-Mills
theory in $6D$ has a dimensional coupling constant and is therefore
non-renormalizable. The ultraviolet behavior of such a theory has
been studied in frame of different approaches
\cite{Fradkin:1982kf,Marcus:1983bd,Marcus:1984ei,Kazakov:2002jd,Buchbinder:2019cnh,
Howe:1983jm,Howe:2002ui,Bossard:2009sy,Bossard:2009mn,Bossard:2015dva,
Buchbinder:2016url,Buchbinder:2017ozh,Bork:2015zaa}. In compare with
the gauge theory with a standard kinetic term, the model with higher
derivatives \cite{Ivanov:2005qf} includes the dimensionless couple
constant and, hence, is renormalizable with respect to the power
counting. This model possesses conformal invariance
\cite{Ivanov:2005kz} at the classical level, which is broken by the
quantum corrections \cite{Smilga:2004cy,Smilga:2006ax}. The one-loop
beta function for this theory was calculated earlier
\cite{Ivanov:2005qf,Casarin:2019aqw,Buchbinder:2020tnc,Buchbinder:2020ovf}.

The work is organized as follows. In section 1, we present the main
conventions and notations which we use in the work. In Section 2, we
develop the quantization procedure for the theory with higher
derivatives in  $6D, \cN = (1,0)$ harmonic superspace using the
background superfield method and discuss the structure of the
one-loop divergences. In Section 3, we consider the structure of the
one-loop divergent contributions in $6D, \cN=(1,0)$ model of
interacting gauge multiplet and hypermultiplet both including high
derivatives. The last section contains a summary of our results and
a discussion of possible directions for future work.

\section{Notations and conventions}
We use notations and conventions from the works
\cite{Buchbinder:2020ovf,Buchbinder:2020tnc}. The $6D, \cN=(1,0)$
harmonic superspace is parameterized by the coordinates $(z, u) =
(x^M, \theta^a_i,u^{\pm i})$, where $x^M$, $M= 0,..,5$, are $6D$
Minkowski space-time coordinates, $\theta^a_i$ with $a=1,..,4\,$,
$i=1,2\,$, stand for  Grassmann variables. The extra bosonic
coordinates $u^{\pm}_i$, $u^{+i}u^-_i =1$ (called harmonics)
parameterize the coset $SU(2)/U(1)$ of the automorphisms group
$Sp(1) \sim SU(2)$ \cite{Howe:1985ar,Zupnik:1986da}. The analytic
coordinates $\zeta = (x_{\cal A}^M, \theta^{\pm a})$ are defined as
 \be
x^M_{\cal A} \equiv x^M + \frac i2
\theta^{+a}(\gamma^M)_{ab}\theta^{-b},\qquad  \theta^{\pm a}=
u^\pm_k\theta^{ak},
 \ee
where $(\gamma^M)_{ab}$ are the antisymmetric $6D$ Weyl
$\gamma$-matrices, $(\gamma^M)_{ab} = - (\gamma^M)_{ba}\,,
(\widetilde{\gamma}^M)^{ab} =
\frac12\varepsilon^{abcd}(\gamma^M)_{cd}\,,$ with
$\varepsilon^{abcd}$ being the totally antisymmetric Levi-Civita
tensor.

We introduce the harmonic spinor covariant derivatives $D_a^+ =
u^+_i D^i_a$ and $D_a^- = u^-_i D^i_a$, which are together with
harmonic derivatives
 \bea
D^{\pm\pm}= u^{\pm i} \frac{\partial}{\partial u^{\mp i}},\qquad
\qquad D^0 = u^{+i} \frac {\partial}{\partial u^{+i}} - u^{-i} \frac
{\partial}{ \partial u^{-i}}.
 \eea
satisfying the algebra
 \bea
    \{D^+_a,D^-_b\}=i(\gamma^M)_{ab}\partial_M\,, \qquad [D^{++}, D^{--}] = D^0,
    \qquad [D^{\pm\pm},D^{\pm}_a]=0\,, \qquad
    [D^{\pm\pm},D^{\mp}_a]=D^\pm_a\,.
 \eea
The integration measures over the full harmonic and analytic
superspaces are defined as follows
 \be
d^{14}z \equiv d^6x_{\cal A}\,(D^-)^4(D^+)^4,\quad d\zeta^{(-4)}
\equiv d^6x_{\cal A}\,du\,(D^-)^4 ,
 \ee
 \be
(D^{\pm})^4 = -\frac{1}{24} \varepsilon^{abcd} D^\pm_a D^\pm_b
D^\pm_c D^\pm_d\,.
 \ee

Besides the analytic gauge superfield $V^{++}$ it is necessary to
introduce a non-analytic harmonic connection $V^{--}$ as a solution
of the harmonic zero-curvature condition \cite{Galperin:2001uw}
 \bea
D^{++} V^{--} - D^{--}V^{++} + i[V^{++},V^{--}]=0\,.  \label{zeroc}
 \eea
Using the superfields $V^{++}$ and $V^{--}$, we construct the gauge
covariant harmonic derivatives $\nb^{\pm\pm}= D^{\pm\pm} + i
V^{\pm\pm}$ and the spinor gauge covariant derivatives
\cite{Ivanov:2005qf}
 \bea
\nb^+_a = D^+_a, \qquad \nb^-_a = D^-_a + i {\cal A}^-_a, \qquad
\nb_{ab} = \pr_{ab} + i {\cal A}_{ab}\,. \label{deriv}
 \eea
Here we have introduced $\nb_{ab} = \f12(\g^M)_{ab} \nb_M$,
$\nb_M=\partial_M-iA_M\,$ and the superfield connections
 \be
{\cal A}^-_a= i D^+_a V^{--}\,,\quad  {\cal A}_{ab} = \f12  D^+_a
D^+_b V^{--}. \label{OtherConn}
 \ee
The covariant derivatives \eq{deriv} satisfy the algebra
\begin{equation}
\{\nb_a^+,\nb^-_b\}=2i\nb_{ab}\,,\qquad
[\nb_c^\pm,\nb_{ab}]=\frac{i}2\ve_{abcd}W^{\pm\, d},\qquad [\nb_M,
\nb_N] = i F_{MN}\,,\label{alg2}
\end{equation}
where $W^{a\,+}$ is the superfield strength of the gauge
supermultiplet,
 \bea
W^{+a}= -\frac{i}{6}\varepsilon^{abcd}D^+_b D^+_c D^+_d V^{--},
\qquad W^{-a} = \nb^{--}W^{+a}\,. \label{Wstr}
 \eea

\section{Four-derivatives $6D$, $\cN=(1,0)$ SYM models}
\paragraph{Model}
We consider the superfield action
 \begin{eqnarray}
 S_0[V^{++}] &=&
 \f{1}{2 g_0^2}\tr \int d\zeta^{(-4)} (F^{++})^2 \nn \\
&& +\frac{1}{f_0^2}\sum\limits^{\infty}_{n=2} \frac{(-i)^{n}}{n} \tr
\int \frac{ d^{14}z\, du_1\ldots du_n }{(u^+_1 u^+_2)\ldots (u^+_n
u^+_1)} V^{++}(z,u_1 ) \ldots V^{++}(z,u_n)\,, \label{S0}
\end{eqnarray}
where $g_0$ is a dimensionless couple constant and $f_0$ is a second
one with the negative dimension. The action \eq{S0} describe
$6D,\,\cN=(1,0)$ SYM model with a four and two derivatives kinetic
terms for Yang-Mills vector field in the component
\cite{Ivanov:2005qf}.  Also we have introduced the superfield
$F^{++} = \tfrac14 D^+_a W^{+a}$, which is an analytic by
construction.

The action \eq{S0} is invariant under the gauge transformation
 \bea\label{tr1}
 \d_\lambda V^{++} = -D^{++} \lambda -
 i [V^{++}, \lambda]\,,\qquad \d_\lambda F^{++} = i[\lambda,F^{++}]\,,
 \label{gtrans}
 \eea
with the Hermitian analytic superfield parameter $\lambda$ taking
values in the Lie algebra of the gauge group.

\paragraph{Background-field method}
First, we split the superfield $V^{++}$ into the sum of classical
'background' superfield $V^{++}$ and 'quantum' one $v^{++}$
 \be
 V^{++}\to V^{++} + g_0v^{++}.
 \ee
The total quantum action after that reads
 \bea
 S_{\rm quant} &=&
S_{0}[V^{++}+g_0v^{++}] + S_{\rm gf}[v^{++}, V^{++}] + S_{\rm gh}.
 \eea
where the  action for ghosts includes Fadeev-Popov and
Nilsen-Kallosh ones
 \bea\label{FP}
S_{FP} &=&\tr\int d\zeta^{(-4)}\,
{\bf b}\nb^{++}(\nb^{++}{\bf c} +[v^{++},{\bf c}]), \\
S_{NK} &=& \frac{1}{2}\tr \int d\zeta^{(-4)}du\, \varphi
(\nb^{++})^2\varphi\,. \label{NK}
 \eea

One can obtain the gauge-fixing term obtained as a result of the
Faddeev-Popov quantization procedure\footnote{A detailed calculation
of the gauge-fixing term reduced to the this form could be found in
\cite{Buchbinder:2020tnc}.} reads
  \bea
S_{GF}[v^{++}, V^{++}] &=& \frac{1}{2\xi}\tr \int d^{14}z du\,
v_{\tau}^{++}
\sB{}^2 v_{\tau}^{++} \nn \\
&& -\frac{1}{2\xi}\tr \int d^{14}z
\f{du_1du_2}{(u^+_1u^+_2)^2}\Big\{v_{\tau,1}^{++}(\sB
v^{++})_{\tau,2} + \frac 12
v^{++}_{\tau,1}\nb^{--}_2[F^{++},v^{++}]_{\tau,2} \Big\},
\label{SGF}
 \eea
where $\xi $ is the arbitrary real parameter.

To calculate the one-loop effective action for the theory under
consideration we decompose the total action up to the second order
over quantum superfields.  Explicit expression for quadratic part of
the action $S^{(2)}$ has the form
  \bea
 S^{(2)} &=& \frac{1}{2\xi}\tr \int d\zeta^{(-4)}\, v_{\tau}^{++}\sB{}^2 v_{\tau}^{++}
 -\f12  \Big(1+\f1\xi\Big)\tr \int d^{14}z \f{du_1du_2}{(u^+_1u^+_2)^2}
 v^{++}_{\tau,1}[F^{++},\nb^{--} v^{++}]_{\tau,2}\nn \\
 && +\f{1}{2}\Big(1-\f1\xi\Big)\tr \int d^{14}z \f{du_1du_2}{(u^+_1u^+_2)^2}
 \Big\{v_{\tau,1}^{++}(\sB v^{++})_{\tau,2} + \frac 12
v^{++}_{\tau,1}[(\nb^{--}F^{++}),v^{++}]_{\tau,2} \Big\}
 \nn \\
 && +\f{g_0^2}{2f_0^2} \int d^{14}z \f{du_1du_2}{(u^+_1u^+_2)^2} v_{\tau,1}^{++}v_{\tau,2}^{++}
 \nn \\
 && + \tr \int d\zeta^{(-4)}\,{\bf b}(\nb^{++})^{2}{\bf c}
  +\frac{1}{2}\tr \int d\zeta^{(-4)}\,\varphi(\nb^{++})^{2}\varphi\,.
  \label{S2}
 \eea
In what follows, we take $\xi=1$ in Eq. \eqref{S2} and rewrite the
terms including $F^{++}\nb^{--}$ integrating it by parts.
  \bea
 S^{(2)} &=& \frac{1}{2\xi}\tr \int d\zeta^{(-4)}\, v_{\tau}^{++}\sB{}^2 v_{\tau}^{++}
 +i \tr \int d^{14}z \f{du_1du_2}{(u^+_1u^+_2)^2} v^{++}_{\tau,1}[({\nb}^{--}{F}^{++}),v^{++}]_{\tau,2}
 \Big\} \nn \\
 && -i\tr \int d^{14}z du_1 du_2\frac{(u_1^+ u_2^-)}{(u^+_1u^+_2)^3}
 v^{++}_{\tau,1}[{F}^{++}, v^{++}]_{\tau,2}\nn \\
 && +\f{g_0^2}{2f_0^2} \int d^{14}z \f{du_1du_2}{(u^+_1u^+_2)^2}
 v_{\tau,1}^{++}v_{\tau,2}^{++}
 \nn
 \\
 && + \tr \int d\zeta^{(-4)}\,{\bf b}(\nb^{++})^{2}{\bf c}
  +\frac{1}{2}\tr \int d\zeta^{(-4)}\,\varphi(\nb^{++})^{2}\varphi\,.
  \label{S22}
 \eea

One can see that the action \eq{S2} contains additional contribution
with both $g_0$ and $f_0$ coupled constant. Our aim is to study the
possible divergent contributions from such a term.

\paragraph{One-loop divergences}
After integration \eq{S22} over quantum superfields we obtain the
following expression for the one-loop effective action
   \bea
\bar{\Gamma}^{(1)}[V^{++}]&=&
 \frac{i}{2}\mbox{Tr}_{(2,2)} \ln \bigg\{
(\sB_1)^2\, (D_1^+)^4\delta^{(-2,2)}(u_1,u_2)
\qquad\nonumber\\
&& + \frac{(D_1^+)^4 (D_2^+)^4}{(u_1^+ u_2^+)^2} \Big[
\frac{g_0^2}{2f_0^2} + i (\nabla^{--}F^{++}) -
2\frac{i(u_1^+ u_2^-)}{(u_1^+ u_2^+)}{F}^{++} \Big]_2 \delta^{14}(z_1-z_2)\bigg\},\qquad \nn \\
&&-i\mbox{Tr}_{(4,0)}\ln\sB  -i\mbox{Tr}\ln \nb^{++}_{Adj}\,.
\label{Gamma0}
 \eea
where $\d^{14}(z_1-z_2)=\d^8(\theta_1-\theta_2)\d^6(x_1-x_2)$. We
are interesting in the divergent contributions to the effective
action \eq{Gamma0} hence for the following analysis we rewrite it as
the sum of two terms,
\begin{eqnarray}\label{Two_Logarithms}
\bar{\Gamma}^{(1)}[V^{++}]&=&
 \frac{i}{2}\mbox{Tr}_{(2,2)} \ln (\sB_1)^2 -i\mbox{Tr}_{(4,0)}\ln\sB  -i\mbox{Tr}\ln \nb^{++}_{Adj}
   \\
&&
 +\frac{i}{2}\mbox{Tr}_{(2,2)}\ln\bigg\{1
 + \frac{1}{(\sB_1)^2}\frac{(D_1^+)^4 (D_2^+)^4}{(u_1^+ u_2^+)^2}
\Big[ \frac{g_0^2}{2f_0^2} + i (\nabla^{--}F^{++})_2 -
2\frac{i(u_1^+ u_2^-)}{(u_1^+ u_2^+)}{F}^{++}_2 \Big] \bigg\}. \nn
\end{eqnarray}
According to \cite{Buchbinder:2017ozh, Buchbinder:2020tnc,
Buchbinder:2020ovf} the first two terms in \eqref{Two_Logarithms} do
not contribute to the divergent part of the effective action. To
calculate the divergent part of the expression \eq{Two_Logarithms},
we need to decompose the logarithm in the last term up to a linear
contributions only. Then we obtain
\begin{eqnarray}
\Gamma_{\rm div}^{(1)}[{V}^{++}] = \Gamma_1 + \Gamma_2 + \Gamma_3 +
\Gamma_4,
\end{eqnarray}
where
\begin{eqnarray}\label{Gamma1}
&& \Gamma_1 = \tr \int d\zeta^{(-4)}_1 \frac{(D_1^+)^4 (D_2^+)^4}{(\sB_1)^2} \frac{(u_1^+ u_2^-)}{(u_1^+ u_2^+)^3} {F}^{++}_2 \delta^{14}(z_1-z_2)\Big|_{2\to 1},\qquad\\
\label{Gamma2} && \Gamma_2 = -\frac{1}{2} \tr \int d\zeta^{(-4)}_1
\frac{(D_1^+)^4 (D_2^+)^4}{(\sB_1)^2}
\frac{(\nabla^{--} F^{++})_2}{(u_1^+ u_2^+)^2}  \,\delta^{14}(z_1-z_2)\Big|_{2\to 1},\qquad\\
\label{Gamma3} && \Gamma_3 = \frac{i g_0^2}{4f_0^2} \tr \int
d\zeta^{(-4)}_1 \frac{(D_1^+)^4(D_2^+)^4 }{(\sB_1)^2}
\frac{\delta^{14}(z_1-z_2)}{(u_1^+ u_2^+)^2}
\Big|_{2\to 1},\\
\label{Gamma4} && \Gamma_4 = -i\Tr\ln
\nb^{++}\vphantom{\frac{1}{2}},
\end{eqnarray}
and $\tr$ stands for the usual matrix trace. The contributions
$\Gamma_1, \Gamma_2$ and $\Gamma_4$ were considered earlier in the
works \cite{Buchbinder:2020tnc,Buchbinder:2020ovf}. The presence of
to the additional contributions $\Gamma_3$ is caused by the standard
two-derivative SYM kinetic term in the classical action of the model
\eqref{S0}.

As the first step of the calculation, we consider the divergent
contribution coming from \eq{Gamma1},
 \be
\Gamma_{1}=  \int d\zeta_1^{(-4)} du_1\, ((\sB_1){}^{-2})^{IJ}
(D^+_1)^4 (D^+_2)^4\frac{(u_1^+ u_2^-)}{(u_1^+u_2^+)^3}
(F^{++}_2)^{JI} \d^{14}(z_1-z_2)\Big|_{2\to 1}\,. \label{Gamma1_1}
 \ee
In this expression, we should comute the operator $(\sB)^{-2}$ to
act on the delta-function. Acting on analytic superfields, the
covariant d'Alembertian reduced to
$$ \sB{}^{IJ} =
\partial^2 \d^{IJ} + i({F}^{++})^{IJ} D^{--} + \dots\,,
$$
where $(F^{++})^{IJ} = - if^{KIJ} {F}^{++K}$ and $f^{KIJ}$ are fully
antisymmetric structure contant, $[t^I,t^J]=if^{IJK}t^K$, with the
standard normalization condition for generators of the gauge group
$\tr( t^I t^J )= \tfrac12 \d^{IJ}$. The logarithmically divergent
contribution in \eq{Gamma1_1} is proportional to the third power of
the inverse operator $\partial^2 =
\partial^M \partial_M$  acting on the space-time
delta-function $\d^6(x_1~-~x_2)$,
 \bea
 \f{1}{(\pr^2)^3} \delta^6(x_1-x_2)\Big|_{2\to 1} =
 \f{i}{(4\pi)^3 \varepsilon}\,, \quad \varepsilon\to 0\,.
 \label{div}
 \eea
Hence, decomposing the inverse d'Alembertian in the series over
$F^{++}$ we are interesting in the terms, which have the
corresponding power over $1/\partial^2$. After calculation we obtain
 \bea
\Gamma_{1,\,\infty} = - \f{4 C_2}{(4\pi)^3 \varepsilon}
 \tr \int d\zeta^{(-4)} du\, ({F}^{++})^2\,.
 \label{Gamma1_Infty}
 \eea
The same calculation can be provided for the contributions
$\Gamma_{2,\,\infty}$ \cite{Buchbinder:2020tnc}. We have
 \be
\label{Gamma2_infty} \Gamma_{2,\,\infty} = 0.
 \ee

Now, let us consider the contribution from the $\Gamma_3$. First, we
calculate the coincident Grassmann points limit $\theta_2\to
\theta_1$ using the identity
 \bea
 (D_1^+)^4 (D_2^+)^4 \delta^8(\theta_1-\theta_2) = (u^+_1 u^+_2)^4
 (D_1^+)^4 (D_1^-)^4 \delta^8(\theta_1-\theta_2).
 \label{Id}
 \eea
and the property
$(D_1^+)^{4}(D_1^-)^{4}\d^{8}(\theta_1-\theta_2)|_{2\to 1} = 1$. One
obtains
 \bea
\label{Gamma3} && \Gamma_{3,\infty} = \frac{i g_0^2}{4f_0^2} \tr
\int d\zeta^{(-4)}_1 \frac{1}{(\sB_1)^2} (u_1^+ u_2^+)^2
\delta^{6}(x_1-x_2)\Big|_{2\to 1,\, \rm{div}}.
 \eea
Calculating the coincident harmonic points limit we need to collect
two harmonic derivatives $\nb^{--}$ from the inverse d'Alembertian.
Indeed, due to the property $D_1^{--}(u^+_1 u^+_2)|_{2\to 1} = - 1$
the first non-vanishing contribution has a structure
 \bea
\frac{1}{(\sB_1)^2} (u_1^+ u_2^+)^2&=& \frac{1}{(\partial^2 +
i{F}^{++} D^{--} + \dots)^2} (u_1^+
u_2^+)^2 \nn \\
&& \approx \frac{1}{(\partial^2)^4}({F}^{++} D^{--})^2 (u_1^+
u_2^+)^2\,.
 \eea
As we see the corresponding contribution has the forth power of the
inverse d'Alembertian and consequently do not contribute to the
divergent part of the effective action. Hence we have
 \be
\label{Gamma3_infty} \Gamma_{3,\,\infty} = 0.
 \ee

The divergent contribution from $\Gamma_4$ in \eq{Gamma4} was
considered earlier in work \cite{Buchbinder:2016url} and has the
form
 \bea\label{Gamma4_Infty}
 \Delta\Gamma_{4,\,\infty}= \f{C_2}{3(4\pi)^3 \varepsilon}\,
 \tr\int d\zeta^{(-4)} du\, ({F}^{++})^2\,.
 \eea

As a result for the divergent contribution of the one-loop effective
action \eq{Gamma0} we sum the non-vanishing contributions
\eq{Gamma1_Infty} and \eq{Gamma4_Infty},
  \be
 \Delta\Gamma^{(1)}_{\rm \infty} = \sum_{i=1}^4\Gamma_{i,\,\infty}
 = -\f{11}{3} \f{C_2}{(4\pi)^3 \varepsilon} \tr
 \int d\zeta^{(-4)} du\, ({F}^{++})^2\,.
 \label{result1}
 \ee
We see that the result divergent contribution to the effective
action do not include the couple constant $f_0$ and coincide exactly
with the superfield expression obtained earlier in
\cite{Buchbinder:2020ovf, Buchbinder:2020tnc}.

\section{SYM  interacting with high-derivative hypermultiplets}
In the work \cite{Buchbinder:2020tnc}, the divergences of the
effective action for $6D,\,\cN=(1,0)$ model of high-derivative gauge
multiplet interacting with the hypermultiplet without high
derivatives were studied in details. In the present section, we
consider the generalization of the considered model. The superfield
action of $6D,\,\cN=(1,0)$ SYM model with a four space-time
derivatives term for both gauge and hypermultiplet superfields
reads\footnote{One can introduce more general action
 \begin{eqnarray}
 S[V^{++},q^+] &=& S_{0} - \int d\zeta^{(-4)} \widetilde q^+ \nb^{++} {\cal F}({f_0^2 \sB}) q^+\,,
\label{S03}
\end{eqnarray}
where ${\cal F}(f_0^2 \sB)$ is arbitrary analytical function. But
this theory contains higher then four derivatives in the
hypermultiplet sector and is out of our consideration in the present
work.}
 \begin{eqnarray}
 S[V^{++},q^+] &=& S_{0} - \int d\zeta^{(-4)} \widetilde q^+ \nb^{++} (1+{f_0^2 \sB})
 q^+\,,
\label{S03}
\end{eqnarray}
where $S_0$ is introduce in \eq{S0}.  We have to note that harmonic
covariant derivative commute with the d'Alembertian on the space of
analytic superfields $q^+$, $[\nb^{++},\sB]q^+ = 0$.

Let us consider the Green function ${\cal G}^{(1,1)}(1|2)$ under
condition
 \bea
 \nb^{++}_1 \Big(1+f_0^2\sB_1\Big) {\cal G}^{(1,1)}(1|2) = \delta_{\cal A}^{(1,3)}(1|2)\,.
 \eea
The formal solution of the equation above reads
 \bea
 \label{Green}
 {\cal G}^{(1,1)}(1|2) =\Big(1+f_0^2\sB_1\Big)^{-1} G^{(1,1)}(1|2)\,, \nn \\
 G^{(1,1)}(1|2) = \f{(D^+_1)^4 (D^+_2)^4}{\sB} \f{\d^{14}(z_1-z_2)}{(u_1^+u_2^+)}\,.
 \eea
The Green function ${\cal G}^{(1,1)}(1|2)$ is generalization of the
known Green function for hypermultiplet \cite{Galperin:2001uw}
without high derivatives.

\paragraph{One-loop divergences}
Let us consider the contribution to the one-loop effective action
from the action \eq{S03} under the following background-quantum
splitting
 \bea
V^{++}\to V^{++}+g_0 v^{++}\,, \qquad q^+\to q^+\,.
 \eea
The further procedure is similar to the calculation of the one-loop
effective action in the previous section. We integrate over quantum
fields the quadratic part of the total quantum action, including the
classical action with the gauge fixing one and actions for ghosts.
After that one can obtain the one-loop contribution to effective
action for the model \eq{S03}:
  \bea
  \bar{\Gamma}^{(1)}[V^{++},q^+]&=& \bar{\Gamma}_0^{(1)}[V^{++}]
  + i\mbox{Tr}\ln \nb^{++}_{R}\Big(1+f_0^2\sB_1\Big) \,. \label{1loop-3}
 \eea
Following the standard procedure to extruct the divergent
contribution from the last term we vary it with respect to $V^{++}$
 \bea
 \d \Tr \ln \nb^{++}\Big(1+f_0^2\sB_1\Big)\Big|_{\rm div}
 &=& i\Tr\,\d V^{++} \Big(1+f_0^2\sB_1\Big) {\cal G}^{(1,1)} + f_0^2 i\Tr\nb^{++}(\d \sB) {\cal G}^{(1,1)} \nn \\
  &=&  i\Tr\,\d V^{++} G^{(1,1)} +2i\Tr (D^+)^4 \nb^{++} \nb^{--}\d V^{--}  {\cal
  G}^{(1,1)}. \label{var1}
 \eea
The divergent contribution from the first term  \eqref{var1} was
considered earlier  (see the details in \cite{Buchbinder:2016url}).
The result is
 \bea \f{T_R}{3(4\pi)^3
\varepsilon}\,
 \tr\int d\zeta^{(-4)} du\, ({F}^{++})^2.
\eea

Let us demonstrate that the second term in \eqref{var1} do not
contains the UV divergent contributions. First, we use the explicit
expresion for the Green function \eq{Green}
 \bea\label{gm2}
&& 2i \tr \int d\zeta^{(-4)} (D^+)^4 \nb^{++} (\nb^{--}\d V^{--})
 \Big(1+f_0^2\sB_1\Big)^{-1} G^{(1,1)}\bigg|_{2=1} \nn \\
&=& 2i \tr \int d^{14}z u\, \nb^{++} (\nb^{--}\d V^{--})
\Big(1+f_0^2\sB_1\Big)^{-1} \f{(D^+_1)^4 (D^+_2)^4}{\sB_1}
\f{\d^{14}(z_1-z_2)}{(u_1^+u_2^+)^3}\bigg|_{2=1}\,,
 \eea
and pass to the full superspace measure. We have to note that we are
interesting only in the divergent contributions of the effective
action and calculate the coincident-points limit for Grassmann
variables, $\theta_2\to \theta_1$ using the identity \eq{Id}. Then
the expression \eqref{gm2} transforms to
 \bea
 2i \tr \int d^{14}z  \nb^{++} (\nb^{--}\d V^{--}) \Big(1+f_0^2\sB_1\Big)^{-1} \f{1}{\sB_1} (u^+_1u^+_2)\d^{6}(x_1-x_2)
 \bigg|_{2=1, \rm div}\,.
 \eea
The function $\Big((1+f_0^2\sB_1){\sB_1}\Big)^{-1}$, we expand in
the power series up to the lower order over inverse covariant
d'Alambertian
 \bea
 \label{gm3}
&& 2i \tr \int d^{14}z  \nb^{++} (\nb^{--}\d V^{--}) \Bigg(
\f{1}{f_0^2\sB_1^2} (u^+_1u^+_2) - \f{1}{(f_0^2)^2\sB_1^3}
(u^+_1u^+_2) \Bigg)\d^{6}(x_1-x_2)\bigg|_{2=1, \rm div}.
 \eea

First we consider the first term in \eqref{gm3}. To extract non-zero
contributions we have to act the harmonic derivative $D^{--}$ on te
factor $(u^+_1u^+_2)$ in coincident points limit. We have
 \bea
\f{1}{f_0^2\sB_1^2} (u^+_1u^+_2)\d^{6}(x_1-x_2)\bigg|_{2=1, \rm div}
&\approx& \f{2}{f_0^2}\f{F^{++}D^{--}}{(\partial^2)^3} (u^+_1u^+_2)\d^{6}(x_1-x_2)\bigg|_{2=1, \rm div} \nn \\
&=& \f{2i}{f_0^2 (4\pi)^3 \varepsilon} F^{++}, \quad \varepsilon\to
0.
 \eea
The second term produces only finite contributions to the effective
action.  Indeed, we have
 \bea
\f{1}{(f_0^2)^2\sB_1^3} (u^+_1u^+_2)\d^{6}(x_1-x_2)\bigg|_{2=1, \rm
div} &\approx& \f{3}{f_0^2}\f{F^{++}D^{--}}{(\partial^2)^4}
(u^+_1u^+_2)\d^{6}(x_1-x_2)\bigg|_{2=1, \rm div}.
 \eea
The corresponding momentum integral in the last expression do not
contain UV divergences. Hence, it does not contribute to divergent
part of effective action. If we return to Eq. \eq{gm3} we see that
possible divergent contribution from this terms is the total
harmonic derivative due to the property $\nb^{++}F^{++}=0$. So the
second term of the variation \eqref{var1} does not contribute to the
divergent part of effective action.

The result for divergences has the same form as in the case
hypermultiplet without high derivatives \cite{Buchbinder:2020tnc}
  \be
 \Delta\Gamma^{(1)}_{\rm \infty} = \sum_{i=1}^5\Gamma_{i,\,\infty}
 = -\f{11C_2+T_R}{3 (4\pi)^3 \varepsilon} \tr
 \int d\zeta^{(-4)} du\, ({F}^{++})^2\,.
 \label{result3}
 \ee

\section*{Summary}

In the present paper we study  the  structure of one-loop
divergences for the higher-derivative $\cN=(1,0)$ supersymmetric
gauge theory in six dimensions. The theory includes the interacting
gauge multiplet and hypermultiplet in arbitrary representation of
the gauge group. The kinetic terms for superfields include up to the
four derivatives for the bosonic fields kinetic terms in components.
The theory is characterized by two coupling constants: arbitrary SYM
theory, which in six dimensions has negative dimensions, and high
derivative SYM theory, which is dimensionless. A few models were
considered: the high derivatives models with the gauge multiplet
only and the model in which the gauge multiplet is coupled to the
hypermultiplet in arbitrary representation of the gauge group, with
the standard and high derivatives kinetic terms. All models were
formulated in harmonic $6D, \,\cN=(1,0)$ superspace ensuring
manifest $\cN=(1,0)$ supersymmetry. We provide the quantization in
the framework of the background superfield method. We study the
one-loop contributions to gauge invariant and manifestly
supersymmetric quantum effective and find the possible divergent
terms in such an action in the gauge multiplet sector.

Finally, we have to note that the results considered above are
obtained in the gauge multiplet sector only. The study of divergent
contributions in the case of a nontrivial hypermultiplet background
requires a separate consideration, which we plan to discuss in
future works. Also, there is some particular interest to study the
structure of the two-loop divergent contributions and to calculate
the two-loop beta-function in the models under consideration. We
also address these questions to the forthcoming works.

\section*{Acknowledgements}
The authors are grateful to I.L. Buchbinder for valuable
discussions. The work of A.S.B. is supported by BASIS Foundation, No
23-1-5-50-1.

\end{document}